\documentclass[12pt,fleqn]{article}
\usepackage{lscape}
\usepackage{graphicx}
\usepackage{color}
\usepackage{enumerate}
\usepackage{hyperref}
\hypersetup{colorlinks = true, linktocpage = true, bookmarksopen = true, linkcolor = blue, urlcolor=blue, citecolor = blue, allcolors=blue}
\usepackage{authblk}
\usepackage[size=footnotesize]{caption}
\usepackage{times}
\usepackage{graphics,latexsym,psfrag,subfig,mathtools}
\usepackage{amsmath}
\usepackage{amssymb}

\usepackage{epsfig}
\usepackage{color}
\usepackage{float}
\newcommand{\ba}{\begin{array}}
\newcommand{\ea}{\end{array}}

\begin{document}
\newcommand{\be}{\begin{equation}}
\newcommand{\ee}{\end{equation}}
\newcommand{\bc}{\begin{center}}
\newcommand{\ec}{\end{center}}
\newcommand{\bdm}{\begin{displaymath}}
\newcommand{\edm}{\end{displaymath}}
\newcommand{\ds}{\displaystyle}
\newcommand{\p}{\partial}
\newcommand{\INT}{\int\limits}
\newcommand{\SUM}{\sum\limits}
\newcommand{\bfm}[1]{\mbox{\boldmath $ #1 $}}
\newcommand{\cre}[1]{{\color{red}{#1}}}
\newcommand{\cgre}[1]{{\color{green}{#1}}}
\title{\bf
A Langevin dynamics approach \\
for multi-layer mass transfer problems}

\author[1]{Oded Farago}
\author[2]{Giuseppe Pontrelli\footnote{Corresponding author. Email: giuseppe.pontrelli@gmail.com}}
\affil[1]{{\footnotesize Department of Biomedical Engineering, Ben-Gurion
  University of the Negev, Be'er Sheva 85105, Israel}}
\affil[2]{{\footnotesize Istituto per le Applicazioni del Calcolo - CNR, Via dei
  Taurini 19, 00185 Rome, Italy}}

\maketitle
\begin{abstract}
We use Langevin dynamics simulations to study the mass diffusion
problem across two adjacent porous layers of different transport property.  At
the interface between the layers, we impose the Kedem-Katchalsky (KK)
interfacial boundary condition that is well suited in a general
situation. A detailed algorithm for the implementation of the KK
interfacial condition in the Langevin dynamics framework is
presented.  As a case study, we consider a two-layer diffusion model of a
drug-eluting stent.  The simulation results are compared with those
obtained from the solution of the corresponding continuum diffusion
equation, and an excellent agreement is shown.
\end{abstract} 
\vspace*{2ex}
\maketitle

\section{Introduction}
\label{sec:intro}

Multi-layer diffusion problems arise in a number of applications of
heat and mass transfer. Some industrial examples are moisture
diffusion in woven fabric composites \cite{pas}, hydrodynamics of
stratified fluids and geological profiles \cite{gros}, environmental
phenomena such as transport of contaminants, chemicals and gases in
layered porous media \cite{liu}, and chamber-based gas fluxes
\cite{liu2}.  Numerous applications concern the biomedical field and
include, for example, transdermal drug delivery \cite{pon1},
drug-eluting stents \cite{pon2} or brain tumor growth
\cite{man}.  While here we focus on multi-layer diffusion,
   other related concepts such as anomalous diffusion, fractal
  kinetics and non-homogenous layers, have been also studied within
  the context of drug release, see e.g.,~\cite{cop,pip,dok}.

Often, the transported material is initially concentrated in one of
the layers from which it propagates to the others by diffusion.
The rate of transfer across the system in mainly determined by the
diffusion coefficients in each layer. In many practical applications
it is essential to regulate the mass flux between layers by suitable
interface conditions. This can be accomplished, for instance, by
placing a selective barrier between adjacent layers, which induces a
chemical potential gradient at the boundary. Another mean for
controlling the transfer rate are membranes which are essentially very
thin boundary layers with a small diffusion
coefficient~\cite{cussler}. In addition to their role in slowing down
the diffusion rate, membranes are also employed for specific
functions, including separation/purification of gases, vapors,
liquids, selection of ions, or other biological functions. Membranes
are routinely used for medical care and individual protection, such as
wound dressing, dialysis, tissue engineering, and controlled release
of drugs.
Membranes are also used for environmental cleaning and protection,
such as water purification and air filtration. A better understanding
of physical behaviour of membranes as rate-controlling barriers can
greatly improve the efficiency of separation and enhance their
performance \cite{yao}.

In this work, we consider simple models for mass transfer in
multi-layered systems.  We assume that the molecules are transported
across the boundaries by passive diffusion only, i.e., no active
transport process is performed to drive the random motion
of molecules. Passive diffusion continues until enough molecules have
passed from a region of higher to a region of lower concentration, to
make the concentration uniform.
When equilibrium is established, the flux of molecules vanish: the
molecules keep moving, but an equal number of them move into and out
of both layers.  Much work has been done from the analytical and
computational point of view for treating multi-layer diffusion in
continuum mechanics. An important aspect of layered systems is the
matching conditions at the interfaces, where an interface is the
common boundary between two layers. Analytical solutions to such
problems are highly valuable as they provide a great level of insight
into the diffusive dynamics and can be used to benchmark numerical
solutions \cite{mar}. Various methods are available for the analysis
and the solution of such problems \cite {cra,car}: The orthogonal
expansion technique and the Green's function approach \cite{tittle,
  mulho,reid,ramk}, the adjoint solution technique \cite{ozi}, the
Laplace transform method \cite{cra,car,lui,carr,carr2}, and finite
integral transforms \cite{olc,mik,pad}. Integral transform techniques
applied to heat transfer problems was reported in great detail in the
book by \"{O}zi\c{s}ik~\cite{ozi}, where several different
transformations are given depending on the situation.  However, there
are severe numerical instabilities and computational drawbacks that
arise when the number of layers increases~\cite{carr}. Other
  papers demonstrate the complexity of solving diffusion problems with
  a large number of layers, either using eigenfunction expansion for
  somewhat different boundary conditions \cite{Oded1}, or based on the
  Green function approach with biological applications \cite{Oded2}.
  Computational complexity of finite difference schemes is widely
  discussed \cite{Oded3}. \par

Recently, a new computational method for studying diffusion problems
in multi-layer systems has been proposed~\cite{far1,far2}. The method
is based on the well-established notion that Brownian dynamics of
particles can be also described by the Langevin's equation (LE)
\cite{vank}.  Therefore, the particle's probability distribution
function (or, equivalently, the material concentration) can be
computed from an ensemble of statistically-independent single particle
trajectories generated by numerical integration of the corresponding
LE.  Integrating LE within each layer is pretty straightforward, and
there are a number of algorithms (Langevin ``thermostats'') that are
widely used for molecular dynamics simulations at constant
temperature~\cite{bbk,lei,gjf}. The key problem is how to perform the
integration during time-steps where the particle moves between layers,
in a manner ensuring that the imposed interlayer conditions are
satisfied.  In Ref.~\cite{far2}, a set of algorithms for handling the
dynamics across sharp interfaces has been introduced. Here we present
an algorithm that combines many types of interfaces (a sudden change
in diffusivity, a semi-permeable membrane, and an imperfect contact),
with the advantage of treating all these cases with a unified
  physical-based method. The new algorithm is applied for studying a
two-layer model of drug release from a drug eluting stent into the
artery. Excellent agreement is found between the LE computational
results and the semi-analytical solution.

\section{Multi-layer systems: diffusion equation}
\label{sec:multi1}
\setcounter{equation}{0} 

Let us consider a composite medium consisting of a number of layered
slabs. A slab is defined here as a plate that is homogeneous and
isotropic, having a finite thickness, but extends to infinity in the
other two dimensions.  In a typical diffusion problem driven by
concentration gradient, most of the mass dynamics occurs along the
direction normal to the layers. We, therefore, restrict our study to a
simplified one-dimensional model across a multi-layer system.  The
concentration of material in each region, $c_i(x,t)$ ($i=1,\ldots,n$),
is governed by the time-dependent diffusion equation
\begin{equation}
  \frac{\partial c_i}{\partial t}=D_i\frac{\partial^2
    c_i}{\partial x^2},
  \label{eq:diffeq}
\end{equation}
where $D_i$ is the diffusion coefficient in the $i$-th region. The
concentrations in the adjacent regions $i$ and $i+1$ must be matched
at the boundary between them, which is located at $x=L_i$. Two
interfacial boundary conditions (IBCs) must be specified at each
interface. If mass is conserved (no source or sink) at the interface,
then the concentration flux must be continuous
\begin{equation} 
 J_i=-D_i\frac{\partial c_i}{\partial x}=-D_{i+1}\frac{\partial
   c_{i+1}}{\partial x}=J_{i+1} \qquad \mbox{at   } x=L_i,   \qquad  t>0.
 \label{eq:bcflux} 
 \end{equation}

The other IBC to be specified at $x=L_i$ depends on the nature of the
interface. The transport of material can be completely blocked by
placing a perfectly reflecting ($J_i=0$) or perfectly absorbing
($c_i=0$) barriers. Typically, however, we are interested at
intermediate situations where the mass flux is not completely blocked,
but only hindered by interfaces whose aim is to control the rate of
mass transfer across the layers. Here, we consider Kedem-Katchalsky
(KK) IBC that reads~\cite{ked,kem}
\begin{equation}
  J_i=P_i \left(c_i-\sigma_i c_{i+1}\right), \qquad \mbox{at } x=L_i,
  \qquad t>0,
  \label{eq:kk}
\end{equation}
where $P_i$ and $\sigma_i$ are, respectively, the permeability and
partition coefficients of the KK condition. We focus on the KK IBC
(\ref{eq:kk}) because it represents the most general case of an
interface where both a discontinuity in the chemical potential and a
semi-permeable membrane are present, in addition to a possible
discontinuity in the diffusion coefficient. The case without a
membrane corresponds to the limit $P_i\rightarrow\infty$, when the KK
IBC must be replaced with
\begin{equation}
  c_i=\sigma_i c_{i+1}, \qquad \mbox{at  } x=L_i, \qquad  t>0
  \label{eq:bcpartition}
\end{equation}
or, otherwise, the flux diverges at the
interface. Eq.~(\ref{eq:bcpartition}) describes the interfacial
condition at an imperfect contact boundary with partition coefficient
$\sigma_i$ arising from the discontinuity in the chemical potential of
the transported molecules in the adjacent layers~\cite{far2}. In the
special case of Eq.~(\ref{eq:bcpartition}) when $\sigma_i=0$ (or,
$\sigma_i\rightarrow\infty$), we have $c_i=0$ (or, $c_{i+1}=0$), which
describes a perfectly absorbing boundary. A subcase of
(\ref{eq:bcpartition}) is $\sigma_i=1$ (a {\em perfect} contact), when
the concentration exhibits no discontinuity for
$P_i\rightarrow\infty$. However, when $P_i$ is finite in eqn
(\ref{eq:kk}), we expect a concentration jump even for $\sigma_i=1$,
as the KK IBC reduces to
\begin{equation} 
 J_i=P_i \left(c_i-c_{i+1}\right), \qquad \mbox{at } x=L_i, \qquad
 t>0,
 \label{eq:bcmembrane}
\end{equation}
which is the IBC describing the effect of a thin semi-permeable
membrane with permeability $P_i$, but without a chemical potential
jump.  Finally, when $P_i=0$, we recover the condition at a perfectly
reflecting boundary, $J_i=0$.

\section{Multi-layer systems: Langevin equation}
\label{sec:langevin}
\setcounter{equation}{0} 

The method presented in ref.~\cite{far2} is based on the description
of the overdamped Brownian motion of particles via the underdamped LE
\begin{equation}
  m\frac{dv}{dt}=-\alpha(x) v+\beta(t)+f(x),
  \label{eq:langevin}
\end{equation}
where $m$ and $v=dx/dt$ denote, respectively, the mass and velocity of
the diffusing particle. This is Newton equation of motion under the
action of a ``deterministic'' force $f(x)$. The impact of the random
collisions between the Brownian particle and the molecules of the
embedding medium is introduced by two additional forces - (i) a
friction force, $-\alpha(x)v$, and (ii) stochastic Gaussian thermal
noise, $\beta(t)$, with zero mean, $\langle \beta(t)\rangle=0$, and
delta-function auto-correlation,
$\langle\beta(t)\beta(t^{\prime})\rangle=2k_BT
\alpha\left(x\left(t\right)\right)\delta(t-t^{\prime})$,
where $T$ is the temperature and $k_B$ is Boltzmann's
constant~\cite{ris}. The friction coefficient, $\alpha$, in LE
and the diffusion coefficients, $D$, in the corresponding
diffusion equation, satisfy the Einstein's relation~\cite{vank,far3}:
\begin{equation}
  \alpha(x) D(x)=k_BT.
  \label{eq:einstein}
\end{equation}

\begin{figure}[t]
\centering\includegraphics[width=0.7\textwidth]{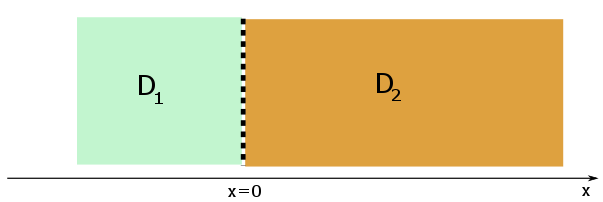}
\caption{The typical two-layer one-dimensional system. A continuous
  flux [see Eq.~(\ref{eq:bcflux1})] is imposed at the interface $x=0$
  (dashed line), together with Kedem-Khatchalsky (KK) condition [see
    Eq.~(\ref{eq:kk1})].}
\label{fig:twolayer}
\end{figure}

In the Langevin dynamics approach to multi-layer diffusion, the
concentration profile, $c(x,t)$, is computed by generating an ensemble
of statistically-independent particle trajectories of duration $t$,
from which a fine-grained histogram can be constructed. We define $c(x,t)$
such that, at $t=0$, the total density is normalized to unity and
essentially represents the initial probability distribution function
of the particles
\begin{equation}
  \int_{-\infty}^{\infty} c(x,0)dx=1.
  \label{eq:normalization}
\end{equation}
The trajectories are calculated by numerically integrating
Eq.~(\ref{eq:langevin}).  To allow for simulations of Langevin
dynamics in multi-layer systems, algorithms were derived
in~\cite{far2} for handling the transition in presence of (i) layers
with different diffusion coefficients,
(ii) a semi-permeable membrane,
and (iii) a step-function chemical potential.
Here, we integrate them into a single unified algorithm for crossing a
KK IBC [Eq.~(\ref{eq:kk})] with continuous flux [IBC
  Eq.~(\ref{eq:bcflux})].  We will not repeat the discussion on the
physical basis underlying the method, but rather present a practical
{\em recipe} describing how to implement the algorithm. To this
purpose, we consider the two-layer system shown in
fig.~\ref{fig:twolayer}, with a step diffusion function
\begin{equation}
  D(x)=\left\{\begin{array}{ll}
  D_1\ \  & x<0  \\
  \label{eq:jumpd}\\
  D_2\ \  & x>0.
  \end{array}\right.
\end{equation}
The continuity of flux $J$ applies at the interface
\begin{equation}
  -D_1\frac{\partial c_1}{\partial x}=-D_2\frac{\partial
   c_2}{\partial x} \qquad \mbox{at   } x=0,   \qquad  t>0,
 \label{eq:bcflux1} 
 \end{equation}
together with the KK IBC
\begin{equation}
  J=P \left(c_1-\sigma c_{2}\right), \qquad \mbox{at } x=0, 
  \qquad t>0.    \label{eq:kk1}
\end{equation}

\subsection{Langevin integrator}

The initial position of the particle is drawn from the
probability distribution $c(x,0)$, Eq. (\ref{eq:normalization}), and the
initial velocity from the Maxwell-Boltzmann distribution 
\begin{equation}
\rho_{\rm
  MB}(v)=\sqrt{ m \over 2\pi k_BT} \exp\left(-{mv^2 \over 2k_BT}\right).
\label{eq:rhomb}
\end{equation}
The trajectory $x(t)$ is then computed by performing discrete-time
integration of LE (\ref{eq:langevin}). For
this purpose, we use the algorithm of Gr{\o}nbech-Jensen and Farago
(GJF)~\cite{gjf}
\begin{eqnarray}
  x^{n+1}&=&x^n+b\,\left[dtv^n+\frac{dt^2}{2m}f^n+\frac{dt}{2m}\beta^{n+1}\right]
  \label{eq:gjfx}\\
  v^{n+1}&=&a\,v^n+\frac{dt}{2m}\left(a\,f^n+f^{n+1}\right)+\frac{b}{m}\beta^{n+1}
  \label{eq:gjfv},
\end{eqnarray} 
to advance the coordinate $x^n=x(t_n)$ and velocity $v^n=v(t_n)$ by
one time step from $t_n=n \, dt$ to $t_{n+1}=t_n+dt$. In the above GJF
equations (\ref{eq:gjfx})-(\ref{eq:gjfv}), $f^n=f(x^n)$, $\beta^n$ is
a Gaussian random number satisfying
\begin{equation}
  \langle\beta^n\rangle=0\ ;\ \langle\beta^n\beta^l\rangle=2\alpha
  k_BTdt\delta_{n,l},
  \label{eq:gjfbeta}
\end{equation}
and the damping coefficients of the algorithm are
\begin{equation}
  b= \ds{1 \over 1+\left(\alpha \, dt/2m \right)},
  \qquad a=b\left[1-\left(\alpha \, dt/2m\right)\right].
  \label{eq:gjfab}
\end{equation}
The GJF integrator is chosen because of its robustness against
discretization time errors, which is critical for achieving accurate
statistics of configurational results. More specifically, it
accomplishes statistical accuracy for configurational sampling of the
Boltzmann distribution in closed systems; and it also provides the
correct Einstein diffusion, $\langle x^2\rangle=2(k_BT/\alpha)t$, of a
freely diffusing particle in an unbounded system with constant
$\alpha$~\cite{gjf,gjf2,gjf3,fink}.

We note that Langevin dynamics is diffusive only on time scales larger
the so called ballistic crossover time $\tau_{\rm
  ballistic}=m/\alpha$, whereas it is predominantly ballistic
(inertial) on much smaller time scales. Generally speaking, the GJF
integrator can be implemented in simulations with relatively large
time steps, $dt>\tau_{\rm ballistic}$, and still produce accurate
statistical results at asymptotically large times~\cite{gjf}. A
criterion for choosing $dt$ can be set by the requirement that the
characteristic variations in $f(x)$, during the time step, should not
be significant, i.e., $|f^{n+1}-f^n|\ll|f^n+f^{n+1}|/2$. This
criterion becomes meaningless when a KK interface is crossed because
the interface exerts a singular, delta-function,
force~\cite{far2}. Nevertheless, we will demonstrate that an accurate
algorithm can be devised provided that the integration is performed in
the inertial regime with $dt\ll\tau_{\rm ballistic}$ (see next
section). This implies that the integration time step in multi-layer
systems is bounded by the ballistic time at the most viscous medium:
\begin{equation}
  dt\ll\tau_{\rm ballistic}^{\rm min}=\frac{m}{{\rm
      max}(\alpha_i)}={\rm
      min}(D_i)\frac{m}{k_BT}.
  \label{eq:taumin}
\end{equation}

\subsection{The case of crossing a discontinuity}
\label{sec:algorithm}

Before presenting the algorithm for crossing a KK type IBC, the
following quantities must be introduced:
\begin{itemize}
  \item The thermal velocity of the particle, which is independent of
  $\alpha$, is given by
  \begin{equation}
  v_{\rm th}=2\int_0^{\infty} v\rho_{\rm MB}(v)
  dx=\sqrt{\frac{2k_BT}{\pi m}},
  \label{eq:v0}
  \end{equation}
where $\rho_{\rm MB}(v)$ is the equilibrium Maxwell-Boltzmann velocity
distribution (\ref{eq:rhomb}).
\item The crossing probability is related to the membrane permeability
  $P$ and to the thermal velocity $v_{\rm th}$ by~\cite{far2}
  \begin{equation}
    p=\frac{2P}{2P+v_{\rm th}}
    \label{eq:pcross}
  \end{equation}
\item At the interface we have a step-function chemical
  potential\footnote{We exclude the limit cases $\sigma=0$ and
    $\sigma\rightarrow\infty$, which correspond to a perfectly
    absorbing IBC. The transition across such an interface is handled
    differently, see section~\ref{sec:stent}.}
  \begin{equation}
    \phi_{\rm step}=k_BT\ln(\sigma)H(x),
    \label{eq:stepphi}
  \end{equation}
  where
\begin{equation}
  H(x)=\left\{\begin{array}{ll}
  0  & x<0 \\
\\
1 & x>0
  \end{array}\right.
\end{equation} 
is the Heaviside step function. The step-function potential result in
a delta-function force $f_{\rm step}=-d\phi_{\rm
  step}/dx=-k_BT\ln(\sigma)\delta(x)$ with a singularity at the
interface. In the proposed computational scheme, the singular
delta-function force is replaced with a sharp, piecewise
constant force 
\begin{equation}
  f(x)=\left\{\begin{array}{ll}
  -\ds\frac{k_BT\ln(\sigma)}{2\Delta_1}\ \
  & -\Delta_1<x<0  \\
\label{eq:thinforce}\\
-\ds\frac{k_BT\ln(\sigma)}{2\Delta_2}\ \ &
0<x<\Delta_2\\ \\ 0\ \ & {\rm elsewhere},
  \end{array}\right.
\end{equation} 
defined in the "small" interval
\begin{equation}
  \left[-\Delta_1, \Delta_2\right]= \left[-\frac{\gamma
      D_1}{v_{th}},\frac{\gamma D_2}{v_{th}}\right],
  \label{eq:bl1}
\end{equation}
with the associated potential
\begin{equation}
  \phi(x)=-\int_{-\infty}^x f(y)dy.
  \label{eq:pwpotential}
\end{equation}
The thickness of  {\em interface layer} (IL) $[-\Delta_1,
    \Delta_2]$
over which the chemical potential changes by $k_BT\ln(\sigma)$ is
controlled by the dimensionless parameter $\gamma$. In the
simulations, $\gamma$ is taken to be of the order of unity such that
$\Delta_i$ ($i=1,2$) is comparable or smaller of the particle mean free path,
$l_{\rm MFP}=2D_i/v_{\rm th}$, i.e. the characteristic distance
traveled by the particle within the ballistic time $\tau_{\rm
  ballistic}$. The condition (\ref{eq:taumin}) guarantees that the
discrete-time trajectory does not hop from side to side of the
interface, but rather passes across the IL and experiences the
influence of the force (\ref{eq:thinforce}).
\item We define the weight function
 \begin{equation}
  W(x)=\exp\left[\frac{\phi(x)-\phi_{\rm step}(x)}{k_BT}\right].
  \label{eq:weight}
\end{equation}
 One can easily check that $W(x)=1$ when $f(x)=0$.
\end{itemize}

\vspace{1cm}

\begin{figure}[t]
  \centering\includegraphics[width=0.7\textwidth]{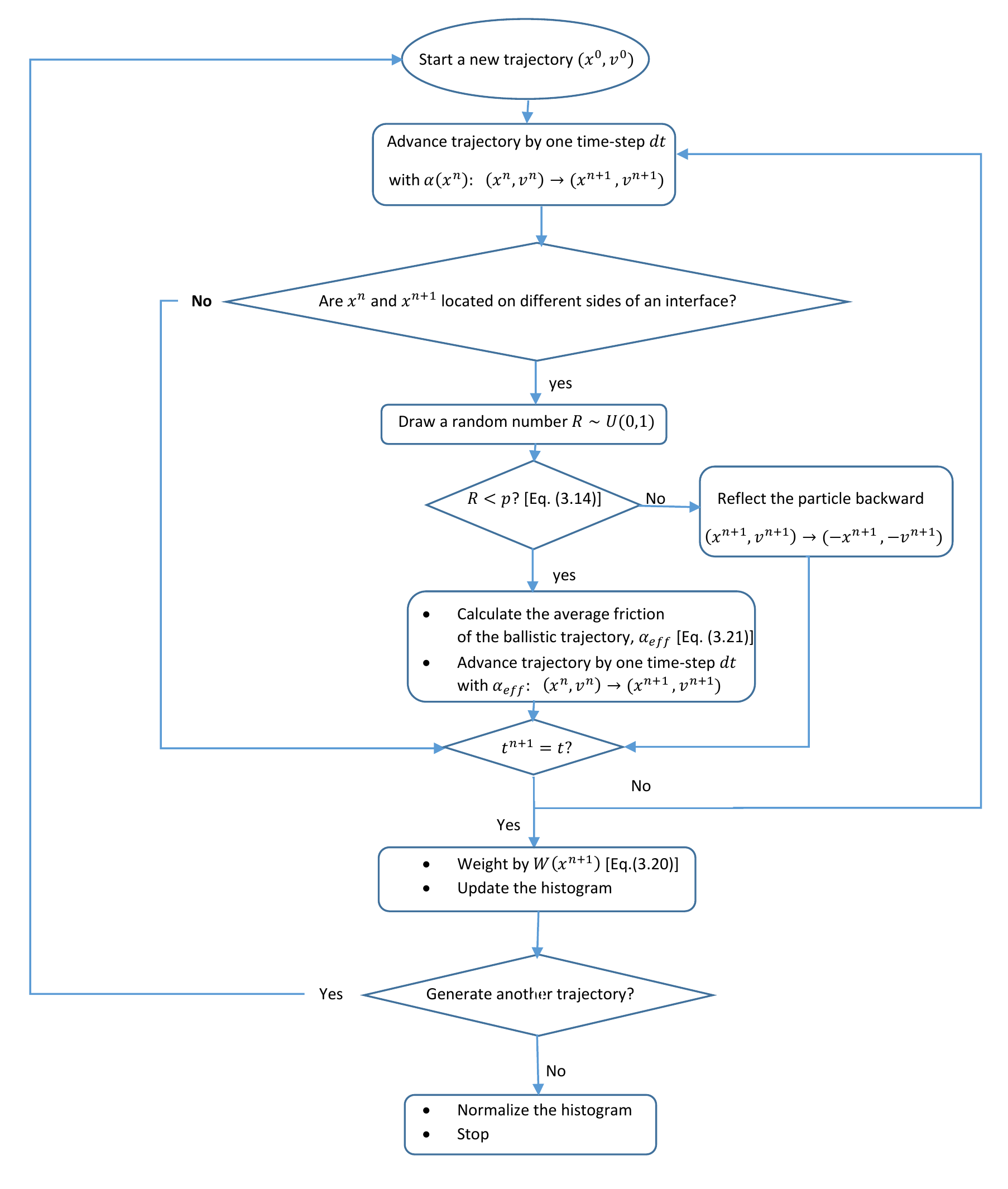}
\caption{A flowchart representation of the algorithm for Langevin
  dynamics simulations of a layered system with a KK interface.}
\label{fig:chart}
\end{figure}

With the above in mind, the algorithm for calculating $c(x,t)$
proceeds as follows:
\begin{enumerate}
  \item Start a new trajectory. Set $t=0$ and $n=0$. Choose the
    initial coordinate $x^0$ from the initial distribution $c(x,0)$,
    and the initial velocity $v^0$ from the equilibrium
    Maxwell-Boltzmann velocity distribution (\ref{eq:rhomb}).
  \item Advance the trajectory from $(x^n,v^n)$ to $(x^{n+1},v^{n+1})$
    by one step $dt$ according to
    Eqs.~(\ref{eq:gjfx})-(\ref{eq:gjfab}), with $f(x)$ given by
    Eq.~(\ref{eq:thinforce}). Use the friction coefficient
    $\alpha(x^n)$ at $x=x^n$.
  \item If $x^{n}$ and $x^{n+1}$ are found on different sides of the
    interface then $x^{n+1}$ needs to be recomputed as follows:
    \begin{itemize}
      \item Choose a random number, ${\cal R}$, uniformly distributed
        between 0 and 1.
      \item If ${\cal R}>p$ [with $p$ given by Eq.~(\ref{eq:pcross})],
        reflect the particle back to the layer from which it arrived
        and set $(x^{n+1},v^{n+1})\rightarrow(-x^{n+1},-v^{n+1})$
      \item If ${\cal R}<p$, allow the particle to move to the
        adjacent layer, and determines $x^{n+1}$ as follows:
        \begin{enumerate}
        \item Calculate the ballistic position $x^{n+1}_{b}=x^n+v^ndt$
        \item Calculate the effective friction coefficient
          \begin{equation}
          \alpha_{\rm
            eff}=\frac{\alpha\left(x^n\right)\left|x^n\right|
            +\alpha\left(x^{n+1}_b\right)\left|x^{n+1}_b\right|}
                {\left|x^n\right|+\left|x^{n+1}_b\right|}
                \label{eq:alphab}
          \end{equation}          
        \item Advance the trajectory from $(x^n,v^n)$ to
          $(x^{n+1},v^{n+1})$ by one step $dt$ according to
          Eqs.~(\ref{eq:gjfx})-(\ref{eq:gjfab}), with the effective
          friction coefficient $\alpha_{\rm eff}$
          (\ref{eq:alphab}). Notice that in some rare cases, the new
          position $x^{n+1}$ will be found on the same side as $x^n$,
          but this is acceptable since small discretization errors are
          always present when encountering a step function diffusion
          function.
        \end{enumerate}
    \end{itemize}    
  \item If $t^{n+1}=t$ then
    \begin{itemize}
    \item Stop the trajectory at $x=x^{n+1}$.
    \item Weight it with the weight function $W(x)$ (\ref{eq:weight}),
      and update the histogram\footnote{In the histogram representation, ${\rm
          hist}_w(x)$, data accumulate in discrete bins. The
        continuous distribution $w(x,t)$ is defined as the total value
        stored within the relevant bin, divided by the bin size.}, ${\rm hist}_w(x)$, for the
      distribution function $w(x,t)$: ${\rm hist}_w(x)={\rm
        hist}_w(x)+W(x)$. 
      \item Return to step 1 if you want to generate another
        trajectory; otherwise go to step 6.
    \end{itemize}
  \item Return to step 2.
  \item Normalize the distribution, $w(x,t)$, to obtain the
    concentration profile, $c(x,t)$:
    \begin{equation}
      c(x,t)=\frac{w(x,t)}{\int_{-\infty}^{\infty}w(x,t)\,dx}.
      \label{eq:histog}
    \end{equation}
\end{enumerate}
Figure~\ref{fig:chart} shows a summary of the algorithm in the form
of a flowchart.

\section{A worked example: a two-layer model of a drug-eluting stent}
\label{sec:stent}
\setcounter{equation}{0}

\begin{figure}[t]
  \centering\includegraphics[width=0.6\textwidth]{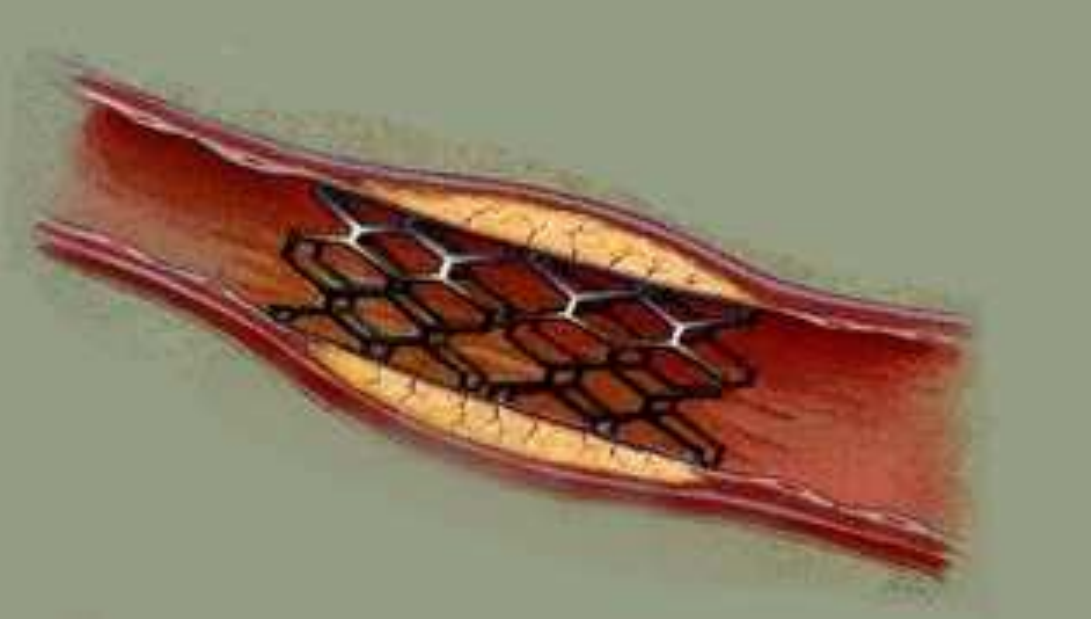}
\caption{A drug-eluting stent implanted in an artery.}
\label{fig:stent}
\end{figure}

In this section we consider a biomedical example where the previous
concepts and algorithms are applied to a simple model of a
drug-eluting stent (DES). Stents are small mesh tubes inserted to keep
open stenosed arteries (see fig.~\ref{fig:stent}). Drug-eluting stents
(DES) also have an additional thin layer of polymer coating the mesh
and eluting a drug.  More precisely, a DES is constituted by metallic
prosthesis ({\em strut}) implanted into the arterial wall and coated
with a thin layer of biocompatible polymer that encapsulates a
therapeutic drug ({\em coating}). Such a drug, released in a
controlled manner through a permeable membrane ({\em topcoat}), is
aimed at healing the vascular tissues or at preventing a possible
restenosis by virtue of its anti-proliferative action against smooth
muscle cells~\cite{mg1,pic}.

\begin{figure}[h]
  \centering\includegraphics[width=0.6\textwidth]{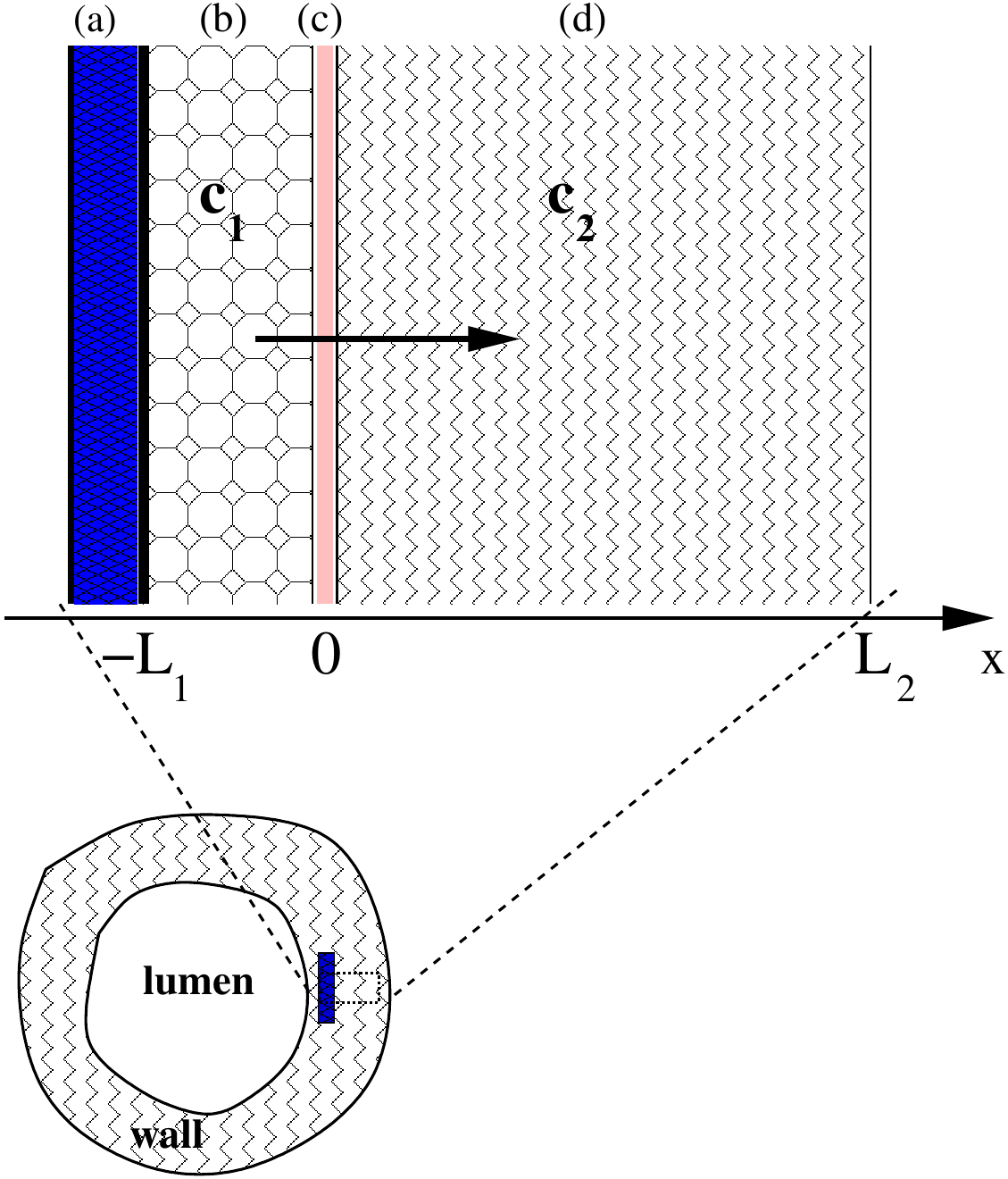}
\caption{Cross-section of a stented artery with the sequence of layers
  for drug dynamics (a) stent strut, (b) coating, (c) topcoat, (d)
  arterial wall (figure not to scale).}
\label{fig:cross}
\end{figure}

To formulate the mathematical problem that serves as a simple DES
model, let us consider a stent coated by a thin layer (of thickness
$L_1$) of polymer containing a drug and embedded into the arterial
wall (of thickness $L_2$), as illustrated in fig.~\ref{fig:cross}. The
complex multi-layered structure of the arterial wall has been
disregarded for simplicity, and a homogeneous material with averaged
diffusion coefficient $D_2$ has been considered.  A small plasma
filtration velocity is present in the wall, but a scaling analysis
shows that this transport effect remains negligible in comparison with
the diffusive one \cite{mg1,pon3}. The diffusion coefficient of the
polymer is $D_1 \ll D_2$. The DES model shown schematically in
fig.~\ref{fig:cross} is a two-layer system similar to the one depicted
in fig.~\ref{fig:twolayer}. The only difference between them is that
here the two layers have a finite extent and two boundary conditions
(BCs) are prescribed to make the mathematical problem well-posed.
Since the strut is impermeable, no mass flux passes through the left
boundary surface, which is modelled by imposing a reflecting boundary
condition: $J_1(-L_1)=0$.  The right side $L_2$, being $L_2 \gg L_1$,
is modeled as an absorbing boundary, namely $c_2(L_2)=0$. At the
initial time ($t = 0$), the drug is contained only in the coating
(layer 1) and it is uniformly distributed at a maximum concentration
$C$:
\begin{align}
&c_1(x,0)=\  C  \qquad \mbox{  for} \quad  -L_1 \leq x\leq 0  \nonumber \\ 
  &c_2(x,0)=\  0\ \ \ \qquad \mbox{   for}  \quad  0  \leq x\leq L_2.
  \label{eq:20layer}
\end{align}
To slow down the drug release rate, a thin membrane (called {\em
  topcoat}) is located at the interface $x=0$ between the two
layers. The topcoat separating the coating and the arterial wall
imposes the KK IBC (\ref{eq:kk1}) between the layers. As no drug is
lost in the topcoat, the continuity of the flux IBC (\ref{eq:bcflux1})
is also assumed.

To summarize, the two-layer diffusion problem is given by the
following set of partial differential equations, with boundary and
initial conditions \cite{pon3}:
\begin{align}
& {\p c_1 \over \p t}- D_1 {\p^2 c_1 \over \p x^2}=0    &\mbox{in}& \,\, 
  [-L_1,0]  \label{sec11} \\
& {\p c_2 \over \p t}  - D_2
{\p^2 c_2 \over \p x^2}=0  &\mbox{in}& \, \, [0 , L_2]  \label{sec8} \\
& -D_1 { \p c_1 \over \p x} = - D_2 { \p c_2 \over \p x}
= P \left( c_1   - \sigma c_2 \right)  
&\mbox{at}& \, \, x=0  \label{sec9} \\
& {\p c_1 \over \p x}=0   &\mbox{at}& \, \, x=-L_1  \label{sec10} \\
&  c_2 =0   &\mbox{at}& \, \, 
x=L_2  \label{sec6} \\ 
& c_1= C, \qquad c_2=0 &\mbox{at}& \, \,     t=0  \label{sec1}
\end{align}
The solution of the above problem is obtained  by separation of variables:
\be
c_i(x,t)=X_i(x) G_i(t)   \qquad\qquad i=1,2 \label{eq1}
\ee
where the spatial functions $X_1$ and $X_2$ satisfy the
 Sturm-Liouville problem:
\begin{align}
&X_1''= -\lambda_1^2 X_1   &\mbox{in}& \, \, [-L_1,0]  \label{sl1}\\
&X_1'=0  &\mbox{at}&  \, \, x=-L_1\label{sl2} \\
&D_1 X'_1= D_2 X'_2 &\mbox{at}& \, \,  x=0 \label{sl3} \\
&&& \nonumber \\ 
&X_2''= -\lambda_2^2 X_2   &\mbox{in}& \, \, [0,L_2] \label{sl4} \\
& X_2=0   &\mbox{at}&  \, \,  x=L_2  \label{sl5} \\
&-D_2 X'_2 + P \sigma X_2= P X_1 &\mbox{at}&  \, \, x=0  \label{sl6}
\end{align}
with:
\be
 D_1 \,\lambda_1^2 = D_2 \, \lambda_2^2   \label{pio}
\ee
The general solution of the ordinary differential eqns. (\ref{sl1}) and (\ref{sl4}) is:
\begin{align}
 X_1(x)=
a_1\cos(\lambda_1 x) + b_1 \sin(\lambda_1 x)  \nonumber \\ 
 X_2(x)= a_2\cos(\lambda_2 x) + b_2 \sin(\lambda_2 x)  \label{sec31}
\end{align}
and 
\be
G_1(t)=G_2(t)=\exp(-D_1 \lambda_1^2 t)=\exp(-D_2 \lambda_2^2 t)
\ee
The eigenvalues $\lambda_i$ and the unknown coefficients $a_i$ and
$b_i$ are computed by 
imposing the BCs and IBCs as follows.
From (\ref{sl2}) and (\ref{sl5}), we have:
\begin{eqnarray}
&&a_1 \sin(\lambda_1 L_1) + b_1 \cos (\lambda_1 L_1)=0  \nonumber  \\
&& a_2 \cos(\lambda_2 L_2) + b_2 \sin (\lambda_2 L_2)=0,    \label{sec42}
\end{eqnarray}  
and from (\ref{sl3}) and (\ref{sl6}), it follows that
\begin{eqnarray}
&& D_1 \:b_1 \:\lambda_1  =  D_2 b_2\: \lambda_2   \nonumber \\
&&- D_2 b_2 \: \lambda_2 + P \sigma  a_2= P a_1. \label{sec62}
\end{eqnarray}  
Eq.~(\ref{sec42})--(\ref{sec62}) form a system of four homogeneous
linear algebraic equations in the four unknowns $a_1, b_1, a_2$ and
$b_2$ .  To get a non trivial solution, it is needed that the
determinant of the coefficient matrix associated with the above system
be equal to zero, that is: \be \tan \left(\sqrt{D_2 \over D_1} L_1
\lambda_2 \right) \left(D_2 \lambda_2 + P \sigma \tan (\lambda_2 L_2)
\right) - \sqrt{D_2 \over D_1}P =0 \label{se11} \ee An infinite
sequence of eigenvalues $\lambda_{21},
\lambda_{22},....\lambda_{2m}... $ is obtained as solutions of the
above transcendental equation (\ref{se11}) (eigencondition).  Hence,
the complete solution of the problem (\ref{sec11})--(\ref{sec1}) is
expressed as a linear superposition of the fundamental solutions:
\begin{align}
&c_1(x,t)=\sum_{m=1}^{\infty} A_m  X_{1m}(x) \: \exp(-D_1 \lambda_{1m}^2 t)   \nonumber \\
&c_2(x,t)=\sum_{m=1}^{\infty} A_m  X_{2m}(x) \: \exp(- D_2 \lambda_{2m}^2 t)
\label{sl33}
\end{align} 
where $A_m$ are determined through the initial conditions  (\ref{sec1})  (see~\cite{pon3} for further details). 

\begin{figure}[t]
\centering\scalebox{1.1}{\includegraphics{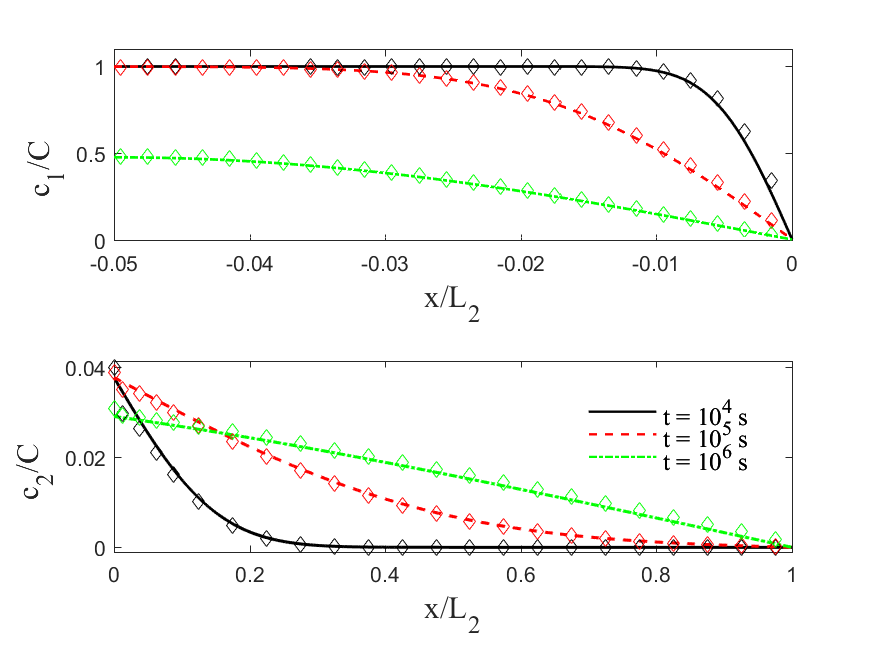}}
\caption{Drug concentration profiles in the coating (above) and in the
  wall (below) for three times (note the different space scales). The
  curves depict the solution obtained by separation of variables
  in~\cite{pon3}, while the markers represent the results of the
  Langevin simulations based on the algorithm described in
  section~\ref{sec:algorithm}.}
\label{fig:results}
\end{figure}

\section{Results}
In the absence of direct experiments, we have chosen the following
parameters which are in the correct range and for which the resulting
release times are consistent with published data \cite{pic,ede,such}:
\begin{align} 
&L_1= 5 \cdot 10^{-4} cm \qquad L_2= 10^{-2} cm \qquad D_1= 10^{-13}
  cm^2/s \qquad D_2=7 \cdot 10^{-11} cm^2/s \nonumber \\ & P= 10^{-5}
  cm/s \qquad \sigma=0.164 \label{par1}
\end{align} 
These parameters, which are representative of the typical scales in
DES, have been chosen based on data in literature for the arterial
wall and heparin drug in the coating layer. The same parameters were
used in ref.~\cite{pon3}, with the exception of $D_1$ and $D_2$ have
been taken $10^3$ smaller, in order to have more
realistic release times. For the Langevin simulations, we use
dimensionless units with $k_BT=1$, $v_{th}=2.523$, $m=10^{-1}$,
$L_1=5$, $L_2=100$, $D_1=10^{-2}$, $D_2=7$. For
  $\gamma=0.5$ in Eq.~(\ref{eq:bl1}), $\Delta_{1}\simeq
  2\cdot10^{-3}\ll L_1$ and $\Delta_{2}=1.387\ll L_2$. In these units,
$P=10^{-1}$. Converting the dimensionless units to physical ones, we
find that $t=1$ in the simulations corresponds to $10^3 s$. The 
time step is set to $5\cdot10^{-5}$, which falls in the ballistic
regime of the Langevin dynamics in both layers, $\tau_{\rm
  ballistic}^{\rm min}=10^{-3}$ [see Eq.~(\ref{eq:taumin})].  We note
that the reflecting boundary at $x=-L_1$ is treated as special cases
of the KK condition with $P=0$ and $\sigma=1$ and is, therefore,
covered by the above algorithm. The absorbing boundary at $x=L_2$
corresponds to $P\rightarrow\infty$ and $\sigma\rightarrow\infty$ (or
$\sigma=0$). In this case, one should assign a very large (or nearly
vanishing) value for $\sigma$ in Eq.~(\ref{eq:thinforce}). In our
simulations we use a simpler approach: We do not introduce a force
near the absorbing interface and, instead, simply terminate and assign
zero weight to each trajectory exceeding $L_2$.

\begin{figure}[t]
\centering\scalebox{0.8}{\includegraphics{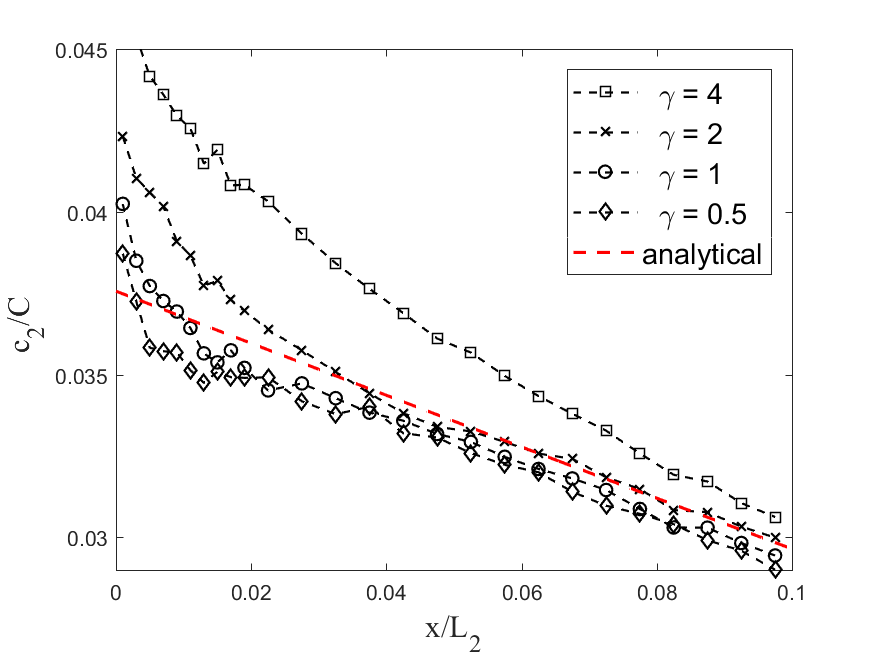}}
\caption {The concentration $c_2$ at $t=10^5$ in the region close
    to the interface. The markers depict the Langevin simulation
    results with different value of $\gamma$ in Eq.~(\ref{eq:bl1}),
    while the dashed red line presents the analytical solution in the
    limit $\gamma\rightarrow 0$, when the chemical potential is given
    by a step-function - see Eq.~(\ref{eq:stepphi}).}
\label{fig:blresults}
\end{figure}

The concentration profiles, $c_1$ and $c_2$, for three values of time
are displayed in fig.~\ref{fig:results}. We observe that the
concentration $c_1$ decays in time, indicating that drug is eluting
from coating to the wall. The concentration at the wall, $c_2$,
increases at short times, and decays at longer times as more and more
drug arrives at the absorbing surface $x=L_2$.  At $x=0$, where the KK
IBC is imposed, we observe a sharp discontinuity in the concentration
that diminishes with time.  The agreement between the semi-analytical
solution (continuous curves) and the Langevin simulation results
(diamond symbols) is excellent, except for deviations near $x=0$ at
the shorter time $t=10^4\ s$. These arise from the approximation of
the delta-function force at the KK interface by the sharp continuous
force (\ref{eq:thinforce}) existing around the interface. The impact
of this approximation on the results are supposedly corrected by the
weight function (\ref{eq:weight}); however, this correction is based
on the ratio of the corresponding Boltzmann factors and, thus, relies
on the assumption that locally the system is at thermal equilibrium
which, strictly speaking, can be only assumed in the overdamped
  limit $\tau_{\rm ballistic}\rightarrow 0$. Fig.~\ref{fig:blresults}
presents results for $c_2$ at $t=10^5$ with larger values of $\gamma$,
zooming in on the region close to the interface. The difference
  between the analytical and numerical solution at $x=0$ provides a
  measure of the computational error, $Er$. Not surprisingly, we find
  that it decreases almost linearly with $\gamma$
  ($\gamma=4:\ Er=0.0095,\ \gamma=2:\ Er=0.0048,\ \gamma=1:\ Er=
  0.0028,\ \gamma=0.5:\ Er=0.0012$)
suggesting that the simulations should be run with the smallest
possible $\gamma$.  Nevertheless, $\gamma$ cannot be reduced
indefinitely since the condition $dt \ll \gamma l_{\rm MFP}/2 v_{\rm
  th}$ is required to ensure that the particle travels within the
IL\footnote{Note that the above condition can be also written as $dt
  \ll \gamma(\pi/2) \tau_{\rm ballistic}$, with $\tau_{\rm ballistic}$
  given by Eq.~(\ref{eq:taumin}), which explains why $\gamma$ should
  be of the order of unity.}. \\

\section{Analysis of discretization errors}

In the last section, we have examined the computational error
  arising from the approximation of a discontinuous chemical potential
  with a sharp piecewise constant jump. Here, we further expand our
  analysis, focusing on the convergence and accuracy of the algorithm
  with respect to the integration time step $dt$. As noted above, we
use the GJF equations (\ref{eq:gjfx})-(\ref{eq:gjfab}) to integrate
the Langevin dynamics, where an ensemble of particle starting on one
side of the interface and spreading across the system. We chose this
integrator because it yields the correct Einstein diffusion, $\langle
x^2\rangle=2\,D \,t=2(k_BT/\alpha)t$, {\em for any time step} when
applied in simulations of a freely diffusing particle. Thus, the
algorithm samples correctly the diffusive dynamics away from the
interface, and discretization errors arise from the segments of the
trajectories when the particle passes close to the interface. These
errors can be minimized by using smaller $dt$, but that would come at
the cost of being able to simulate a smaller number of trajectories
per CPU time, which would increase the statistical noise. In order to
analyze the convergence of the numerical method with respect to $dt$,
we repeat the simulations of a system with IL parameter $\gamma=0.5$
for a sequence of decreasing time steps $dt_k$ ($k=0,1,2,....$). As a
reference case, we set $dt_0=25\cdot10^{-4}$ which is 50 times larger
than the minimal time step $dt$ used to generate the results in
fig.~\ref{fig:results} and 2.5 times larger than the ballistic time,
as computed from Eq.~(\ref{eq:taumin}). We quantify the {\em distance}
between the concentration profiles $c^{(k)}$ corresponding to
subsequent time-steps through the Euclidean norm
  \begin{equation}
    E^{k}( \cdot, t)={\left\| \frac{c^{(k)}- c^{(k-1)}}{c^{(k)}}\right\|_2} ,
    \qquad k=1,\ldots,5
    \label{eq:norm}
  \end{equation}
The results of the analysis are summarized in table~1. The table shows
a clear convergence at smaller time steps and indicates that choosing
$dt=5 \cdot 10^{-5}$ for the simulation results in
fig.~\ref{fig:results} yields a satisfactory accurate solution. The
significant drop in $E^{k}$ between $k=2$ and $k=3$ is probably due to
the fact that $dt_2$ is not sufficiently smaller compared to the
ballistic time ($dt_2=5\cdot10^{-4}=\tau_{\rm ballistic}/2$). Thus,
for the smaller $k$ values in the table the error is predominantly a
systematic discretization one, while for the larger values of $k$ is
dominated by statistical noise.
  
\begin{table}[t]
  \begin{center}
 \label{tab:parameters}
 \begin{tabular}{|c|l|c|c|c|}
   \hline
   Case &  Time step    & $E(\cdot, 10^4)$&  $E (\cdot, 10^5)$ &$E (\cdot, 10^6)$ \\  \hline
    $1$ &  $dt_1=20\,dt$ &    0.0138  &    0.0083  &  0.0080 \\ \hline
    $2$ &  $dt_2=10\,dt$ &    0.0120  &    0.0142  &  0.0386 \\ \hline
    $3$ &  $dt_3=5\,dt$  &    0.0067  &    0.0069  &  0.0074 \\ \hline
    $4$ &  $dt_4=2\,dt$  &    0.0051  &    0.0043  &  0.0085 \\ \hline
    $5$ &  $dt_5=dt$   &    0.0025  &    0.0024  &  0.0057 \\ \hline \hline
 \end{tabular}
 \caption{Norm of the profile difference at three times for a sequence
   of decreasing time steps $dt_k$.}
 \end{center}
\end{table} 
  
To summarize, the simulation results shown in
fig.~\ref{fig:results} represents an acceptable compromise between
accuracy and computational efficiency, dictated by the available
  CPU time, the high aspect ratio ($L_2/L_1=20$), and the
large diffusivity contrast ($D_2/D_1=700$).

\section{Conclusions and perspectives}

We proposed an algorithm for Langevin dynamics simulations in
diffusive multi-layer systems, with flux continuity and KK
interface condition separating regions of different diffusivity. The
proposed method is based on accumulating statistics from a large
number of independent single particle trajectories. These are
  produced by a Langevin dynamics discrete-time integrator, and the
  proposed algorithm describes how the integration is set up when the
  particle crosses an interface. From the ensemble of Langevin
  dynamics trajectories, we generate a fine-grained histogram of the
concentration profile that solves the corresponding continuum
diffusion equation.

To validate the algorithm, we consider the case study of
  two-layer model for a DES that can be solved semi-analytically by
  separation of variables. The agreement between this solution and our
  computational results is shown to be very good. We also use this
  example to assess the accuracy and stability of the method. Our
  analysis suggests that two parameters of the simulations need to be
  carefully chosen: (i) The integration time step that must be smaller
  than the ballistic time of the Langevin dynamics, and (ii) the width
  of the interface layer over which the step-function potential energy
  is approximated. Reducing the values of these parameters improves
  the accuracy of the results, but also increases the computational
  cost since more iterations are needed for generating each
  trajectory. A careful choice should balance between these two
  aspects, and depends on the problem in question and on the available
  computational resources.

  While the example discussed here concerns a two-layer system, it
  should be stressed that a clear advantage of the Langevin dynamics
  algorithm is in dealing with multi-layer systems that have relevance
  to applications in many scientific and engineering disciplines. The
  method can be straightforwardly generalized to any number of
  interfaces, simply by employing the algorithm whenever a trajectory
  encounters one of the interfaces. The simplicity of the algorithm is
  in contrast to analytical solutions that, in general, become
  increasingly complex and computationally inefficient with larger
  number of layers. In a future work we plan to present studies of
  multi-layered systems to demonstrate this important feature of the
  method.

  Another direction is to extend the method to two- and
  three-dimensional composite systems. We also intend to consider
  examples where other mechanisms besides passive diffusion,
  e.g. advection and mass degradation, are included. For the specific
  application of drug-eluting stent considered herein, additional
  efforts are needed to assess and evaluating the relative influence
  of the various factors, including material properties.

\vspace{1cm} 
\noindent{\bf Acknowledgments} \\ Funding from the European Research
Council under the European Unions Horizon 2020 Framework Programme
(No. FP/2014-2020)/ERC Grant Agreement No. 739964 (COPMAT) is
acknowledged.

\end{document}